\documentclass[aps,pra,twocolumn,10pt,showpacs]{revtex4}
\usepackage{amssymb, amsmath}
\usepackage{graphicx,color}
\usepackage{ulem,hyperref}

\begin{document}
\title{Entanglement Witnesses for Indistinguishable Particles}
\author{A. Reusch}\affiliation{Arbeitsgruppe Theoretische Quantenoptik, Institut f\"ur Physik, Universit\"at Rostock, D-18051 Rostock, Germany}
\author{J. Sperling}\email{jan.sperling@uni-rostock.de}\affiliation{Arbeitsgruppe Theoretische Quantenoptik, Institut f\"ur Physik, Universit\"at Rostock, D-18051 Rostock, Germany}
\author{W. Vogel}\affiliation{Arbeitsgruppe Theoretische Quantenoptik, Institut f\"ur Physik, Universit\"at Rostock, D-18051 Rostock, Germany}

\date{\today}

\begin{abstract}
	We study the problem of witnessing entanglement among indistinguishable particles.
	For this purpose, we derive a set of equations which results in necessary and sufficient conditions for probing multipartite entanglement between arbitrary systems of Bosons or Fermions.
	The solution of these equations yields the construction of optimal entanglement witnesses for partial and full entanglement in discrete and continuous variable systems.
	Our approach unifies the verification of entanglement for distinguishable and indistinguishable particles.
	We provide general solutions for certain observables to study quantum entanglement in systems with different quantum statistics in noisy environments.
\end{abstract}

\maketitle

\section{Introduction}

	Nonlocal correlations among many particles or quantized fields are one key element of the quantum nature of physics~\cite{EPR,SCHROE1,SCHROE2,Review1}.
	Applications in metrology use this quantum feature to beat classical limitations~\cite{QM1,QM2,QM3,QM4}.
	Quantum entanglement has been also studied as a resource for quantum information technologies~\cite{QCQI}.
	For a fundamental characterization, a lot of attention has been devoted to verify entanglement between distinguishable particles (DP) or multiple degrees of freedom~\cite{QE1,QP1}.

	Indistinguishable particles (IP), on the other hand, are indispensable for understanding the properties of many-particle quantum systems.
	For IP systems having a certain spin statistics~\cite{SPAIEF}, however, even the notion of entanglement itself has no generally accepted definition~\cite{EIP}.
	For example, the two-Fermion or two-Boson state,
	\begin{align}\label{eq:KeyExample}
	\begin{aligned}
		|\uparrow\rangle\otimes|\downarrow\rangle - |\downarrow\rangle\otimes|\uparrow\rangle \cong& |\uparrow\rangle\wedge|\downarrow\rangle,
		\\|\uparrow\rangle\otimes|\downarrow\rangle + |\downarrow\rangle\otimes|\uparrow\rangle \cong& |\uparrow\rangle\vee|\downarrow\rangle,
	\end{aligned}
	\end{align}
	is a Bell-like entangled state using the tensor product $\otimes$, and, at the same time, it is a product state in the notion of the antisymmetric product $\wedge$ or symmetric product $\vee$, respectively.
	This ambiguity originates from the fact that the (anti)symmetrization requirement of (Fermion)Boson systems has formally the same structure as a nonlocal superposition.
	The left-hand side of Eq.~\eqref{eq:KeyExample} is closely related to the well established theory of entanglement between distinguishable subsystems~\cite{RFW}.
	Hence, we will focus on the right-hand side~\cite{QCTFS,GMW02,GM04}.
	That is, for Fermions and Bosons, the states~\eqref{eq:KeyExample} are separable product states in the exterior algebra and symmetric algebra, respectively.
 	In general, we will focus on the following question: {\it ''How does one certify quantum entanglement which does not rely on (anti)symmetrization?''}

	For bipartite pure states, the relation between entanglement for DP and the tensor product is represented by the Schmidt decomposition~\cite{QCQI}.
	Whenever a single tensor product state is sufficient to expand a pure state, it is separable.
	In analogy, entanglement between IP is characterized by the Slater decomposition~\cite{QCTFS,GMW02,GM04}.
	A pure Fermion or Boson state is separable, if it is a single antisymmetric or symmetric product state, given in terms of Slater determinants or permanents, respectively; see also~\cite{QCSIP,PY01} for a slightly different approach.
	A classical mixture of product states extends the corresponding definitions to mixed quantum states.
	If such a representation is impossible, the state under study is entangled.

	Irregardless of the product for constructing compound Hilbert spaces, $\otimes, \wedge, \vee$, one can detect quantum correlation via so-called entanglement witnesses~\cite{SMS,SMSn,QCTFS}.
	A witness is an observable which is non-negative for all separable states, and may be negative for entangled states.
	Such criteria have been successfully applied to experimentally probe quantum correlations~\cite{BWKGWGHBLS04,VCMM09,DRSJCGFGDS10,JNKGLGCCP10,AGDLRL13,DLTEK14}.
	In the same dimensions, the characterization and application of entanglement in systems of IP gained an increasing importance during the past years; see, e.g.,~\cite{SSN,EoIP,MTE,ERSIP,BESIP,USEIP,EPUA,CMEIP,QQCFS,QCIP,O14,OK13,2015}.
	In particular, the entanglement of multiple qubits, realized in spin systems, has been investigated~\cite{KCL05,ATSL12,PS09,SMLZHPSO14}, and the question of how to extract entanglement for applications from systems  subjected to the Pauli principle has been addressed~\cite{EEIP,EEIPlater}.
	Recently, a method for the construction of optimized multipartite entanglement witnesses for DP has been proposed~\cite{MEW} and applied to perform a full entanglement analysis of experimentally generated multimode states~\cite{Stefan}.

	In the present contribution, we derive equations which allow the construction of optimized, necessary, and sufficient entanglement probes for Boson and Fermion systems.
	The formalism is applicable to arbitrary numbers of particles, partially and fully entangled states, and discrete and continuous variable quantum systems.
	Our method unifies the detection of entanglement for DP and IP.
	Furthermore, we explicitly construct witnesses to demonstrate the strength of this technique to certify entanglement for different spin statistics in noisy environments.

	This work is structured as follows.
	In Sec.~\ref{Sec:SepFermionsBosons}, we discuss the notion of multipartite separable states of IP.
	The construction of corresponding entanglement witnesses is derived in Sec.~\ref{Sec:ConstructWitnesses}.
	Bipartite examples are studied for IP and related to the systems of DP in Sec.~\ref{Sec:BipartiteExample}.
	Section~\ref{Sec:MultipartiteExample} is devoted to establish spin statistics independent witnesses in multimode systems.
	We conclude in Sec.~\ref{Sec:Summary}.

\section{$K$-separable Fermion and Boson states}\label{Sec:SepFermionsBosons}

	The formulation of multipartite entanglement for DP is based on the tensor product structure of compound Hilbert spaces $\mathcal{H}^{\otimes N}$.
	A $N$-partite quantum state $\hat{\sigma}$ is fully separable, if it can be written as a convex combination of product states of the subsystems~\cite{RFW},
	\begin{align}\label{eq:separability}
		\hat{\sigma} = \int dP(a_1,\dots,a_N) \frac{|a_1, \dots , a_N\rangle\langle a_1 , \dots , a_N|}{\langle a_1 , \dots , a_N | a_1 , \dots , a_N\rangle}.
	\end{align}
	Here, $|a_1 , \dots , a_N \rangle=|a_1\rangle\otimes\dots\otimes|a_N\rangle$ are, in general, unnormalized $N$-partite product vectors, and $P$ is a classical probability distribution.

	A fundamental postulate of quantum mechanics for Bosons or Fermions is that the quantum states are symmetric or antisymmetric upon exchange of the subsystems, respectively.
	This restricts the physical states to the (anti)symmetric subspace of the $N$-fold tensor product Hilbert space, $\mathcal{H}^{\wedge N},\mathcal{H}^{\vee N}\subset\mathcal{H}^{\otimes N}$.
	A projection from the tensor product space to these subspaces is given by the permutation operators $\hat\Pi^{\pm}$,
	\begin{align}\label{eq:sym-projector}
		\hat{\Pi}^{\pm} \, |a_1, \dots, a_N\rangle  = \sum_{\sigma \in S_N}\! \frac{(\pm 1)^{|\sigma|}}{N!}|a_{\sigma(1)}, \dots , a_{\sigma(N)}\rangle,
	\end{align}
	where $|\sigma|$ and $(\pm 1)^{|\sigma|}$ denote the parity and the sign of the permutation $\sigma\in S_N$, respectively.
	Other projections might be similarly studied, which allow a generalization to other parastatistics.
	Now, the (anti)symmetric product states can be identified as~\cite{EMSIP}
	\begin{align}
		|a_1\rangle \wedge \cdots \wedge |a_N\rangle &\cong \hat{\Pi}^{-}  \, |a_1 ,  \dots ,  a_N\rangle, \label{eq:antisym_product}\\
		|a_1\rangle \vee \cdots \vee |a_N\rangle &\cong \hat{\Pi}^{+}  \, |a_1 , \dots , a_N\rangle. \label{eq:sym_product}
	\end{align}
	In general, for every vector $|\psi\rangle\in\mathcal H^{\otimes N}$, we get the (anti)symmetric vector in the projected subspace as
	\begin{align}\label{eq:GeneralVector}
		|\psi^\pm\rangle=\hat\Pi^\pm|\psi\rangle.
	\end{align}
	See Appendix~\ref{Sec:Subspaces} for the symmetrization of operators.

	Equations~\eqref{eq:antisym_product} and~\eqref{eq:sym_product} define fully or $N$-separable Fermions and Bosons, respectively.
	More involved is the notion of $K$-separable states; see~\cite{QE1,QP1} for introductions.
	In systems of DP a $K$-separable vector $|\psi_K\rangle\in\mathcal H_N$ is defined as a product vector
	\begin{align}\label{eq:K-product}
		|\psi_K\rangle=|b_1\rangle\otimes\dots\otimes|b_K\rangle=|b_1,\dots,b_K\rangle,
	\end{align}
	for positive integers $(N_1,\dots, N_K)$, $|b_k\rangle\in\mathcal H^{\otimes N_k}$, and $\sum_{k=1}^K N_k=N$.
	The tuple $(N_1,\dots, N_K)$ defines a partitioning of the $N$-fold Hilbert space.
	Further note that $|b_k\rangle$ is, in general, not a product state in $\mathcal H^{\otimes N_k}$. 
	Finally, a $K$-separable (anti)symmetric vector is defined as
	\begin{align}\label{eq:K-SepVector}
		|\psi_K^\pm\rangle=\hat\Pi^\pm|\psi_K\rangle=\hat\Pi^\pm|b_1,\dots,b_K\rangle.
	\end{align}

	Since the permutation is applied in~\eqref{eq:K-SepVector}, the initial ordering of the Hilbert spaces in~\eqref{eq:K-product} does not play a role, i.e., the partition $(N_1,\dots, N_K)$ relates to the Hilbert space sequence $\mathcal H^{\otimes N_1}\otimes\dots\otimes\mathcal H^{\otimes N_K}$ for DP or -- replacing $\otimes$ by $\wedge,\vee$ -- for IP.
	Let us point out that $(N_1,\dots,N_K)$ and $(N'_1,\dots,N'_K)$ define, in general, different partitions if these tuples are not identical up to a permutation of indices.
	For example, the three partition $(2,3,1)$ of a six mode system describes the same partitioning as $(1,2,3)$, but it differs from $(2,2,2)$.

	An explicit example of partial separability of DP is the tripartite state:
	\begin{align}\label{Eq:AppPartSepDP}
	 	|\Psi\rangle=|0\rangle\otimes(|1\rangle\otimes|2\rangle+|3\rangle\otimes|4\rangle)\in\mathcal H^{\otimes 3},
	\end{align}
	using a single-mode orthonormal basis $\{|0\rangle,\ldots,|4\rangle\}$.
	Similarly, one can construct partially separable states of DP in~\eqref{eq:K-SepVector}, e.g.,
	\begin{align}
	\begin{aligned}
		|\Psi^+\rangle\cong& |0\rangle\vee(|1\rangle\vee|2\rangle+|3\rangle\vee|4\rangle)\in\mathcal H^{\vee 3}
		\\\text{ and }
		|\Psi^-\rangle\cong& |0\rangle\wedge(|1\rangle\wedge|2\rangle+|3\rangle\wedge|4\rangle)\in\mathcal H^{\wedge 3}.
	\end{aligned}
	\end{align}
	In Appendix~\ref{Sec:PartSepIP} we prove -- independently from the method to be developed later on -- that these projected states, $|\Psi^\pm\rangle{=}\hat\Pi^{\pm}|\Psi\rangle$, are indeed partially separable, $K=2$, and not fully separable, $K\neq3$.

	Based on the (anti)symmetric product, one gets a general definition of $N$-separability for IP~\cite{EMSIP,OK13}.
	Additionally, a $N$-Fermion or $N$-Boson quantum state $\hat{\sigma}$ is $K$ separable, $1\leq K\leq N$, if it can be written as a convex combination of (anti)symmetric product states,
	\begin{align}\label{eq:separability_identical}
		\hat{\sigma} {=}\! \int\!\! dP(b_1,\dots,b_K) \frac{\hat{\Pi}^{\pm} |b_1, \dots , b_K\rangle\langle b_1 , \dots , b_K|\hat{\Pi}^{\pm}}{\langle b_1, \dots , b_K | \hat{\Pi}^{\pm} |b_1, \dots , b_K\rangle}.
	\end{align}
	Exchanging the tensor product $\otimes$ by either the symmetric product $\vee$ or the antisymmetric product $\wedge$, cf. Eqs.~\eqref{eq:antisym_product} and~\eqref{eq:sym_product}, keeps the separability definition~\eqref{eq:separability} of DP structurally preserved.
	If a state cannot be written according to definition~\eqref{eq:separability_identical} for $K=N$, then entanglement between IP is certified beyond any correlation that can arise from the (anti)symmetrization requirement of the quantum statistics itself.

\section{Construction of entanglement witnesses}\label{Sec:ConstructWitnesses}

\subsection{Separation of entangled states}

	Since Eq.~\eqref{eq:separability_identical} defines a closed, convex set of states, the Hahn-Banach separation theorem is applicable~\cite{Y08,M33}.
	It states that for any closed, convex subset $\mathcal C$ of a Banach space and a point, $x\notin\mathcal C$, there exists a linear and continuous functional $f$ that separates these sets: $f(x)>\sup_{c\in\mathcal C} f(c)$.
	In our case the Banach space is the set of Hermitian trace-class operators and the dual space is isomorphic to the set of bounded operators, i.e., $f(\hat\rho)=\mathrm{tr}(\hat\rho\hat L)$ for some Hermitian $\hat L$.
	This means for the problem under study that for any $K$-entangled state of IP, $\hat\varrho=\hat\Pi^\pm\hat\varrho\hat\Pi^\pm$, exists a bounded Hermitian operator $\hat{L}$, such that
	\begin{align}\label{eq:entanglement_condition_L}
		\mathrm{tr}( \hat{L} \hat{\varrho}) > \sup\{\mathrm{tr}( \hat{L} \hat{\sigma}):\text{for all $\hat{\sigma}$ in Eq.~\eqref{eq:separability_identical}}\}.
	\end{align}
	Related approaches to identify entanglement of DP can be additionally found in Refs.~\cite{SMS,SMSn,SV09,MEW}.
	Since the separation theorem ensures the existence of such an operator, we have a necessary and sufficient condition in terms of observables $\hat L$ probing multipartite entanglement between IP in finite and infinite dimensional spaces.
	However, finding the proper $\hat L$ for a given state $\hat\varrho$ and determining the least upper bound of the right-hand side of inequality~\eqref{eq:entanglement_condition_L} are cumbersome problems.
	In the following, we will propose a method to address the latter aspect.

\subsection{Witness construction through optimization}

	It is sufficient to take the least upper bound on the right-hand side of inequality~\eqref{eq:entanglement_condition_L} over all product vectors, being the extremal points of the given convex set of $K$-separable states.
	Moreover, the operator $\hat{\Pi}^{\pm}$ plays the role of the identity in the (anti)symmetric subspace.
	Combining these facts allows us to write condition~\eqref{eq:entanglement_condition_L} in terms of the expectation value of the entanglement witness operator:
	\begin{align}
		\label{eq:witness_representation_sym} \hat{W}_\pm &=G\hat{\Pi}^{\pm} - \hat{\Pi}^{\pm}\hat{L}\hat{\Pi}^{\pm},\\
		\label{eq:witness_opt} G&=\sup\left\{\frac{\langle b_1,\dots,b_K|\hat\Pi^\pm\hat L\hat\Pi^\pm|b_1,\dots,b_K\rangle}{\langle b_1,\dots,b_K|\hat{\Pi}^{\pm}|b_1,\dots,b_K\rangle}\right\},
	\end{align}
	where the least upper bound is taken over all product vectors~\eqref{eq:K-product}; see also~\cite{QCTFS,QCSIP}.
	Interestingly, the simple modification $\hat{\Pi}^{\pm} \to \hat 1$ in Eqs.~\eqref{eq:witness_representation_sym} and~\eqref{eq:witness_opt} yields the corresponding construction of witnesses for DP~\cite{T05,MEW}.

	The least upper bound in Eq.~\eqref{eq:witness_opt} defines an optimization of the Rayleigh quotient,
	\begin{align}\label{eq:RayleighQ}
		g = \frac{\langle b_1,\dots,b_K|\hat\Pi^\pm\hat L\hat\Pi^\pm |b_1,\dots,b_K\rangle}{\langle b_1,\dots,b_K|\hat{\Pi}^{\pm}|b_1,\dots,b_K\rangle}\to G,
	\end{align}
	under the constraint that the denominator exists, i.e., $\hat\Pi^\pm|b_1,\dots,b_K\rangle\neq 0$.
	The optimization is carried out as a derivative of the Rayleigh quotient:
	\begin{align}\label{eq:derivative_rayleigh}
		0=\frac{\partial g}{\partial \langle b_j|}
		\text{ for $j=1,\dots,K$.}
	\end{align}
	We define the abbreviation $\hat X_{\overline{b_k}}$ for a Hermitian operator $\hat X$, acting on $\mathcal H^{\otimes N}$, by the relation
	\begin{align}
		&\langle x|\hat X_{\overline{b_k}}|y\rangle=\langle b_1,\dots,b_{k-1},x,b_{k+1},\dots,b_K|\hat X\\
		\nonumber &\phantom{\langle x|\hat X_{\overline{b_k}}|y\rangle=}\times|b_1,\dots,b_{k-1},y,b_{k+1},\dots,b_K\rangle,
	\end{align}
	for all $|x\rangle,|y\rangle\in\mathcal H^{\otimes N_k}$.
	Now, the derivative reads as
	\begin{align}
		&0=\frac{\partial g}{\partial \langle b_k|}=\frac{\partial}{\partial \langle b_k|}\frac{\langle b_k|{(\hat\Pi^\pm\hat L\hat\Pi^\pm)}_{\overline{b_k}}|b_k\rangle}{\langle b_k|{(\hat\Pi^\pm)}_{\overline{b_k}}|b_k\rangle}\\
		=&
		\frac{{(\hat\Pi^\pm\hat L\hat\Pi^\pm)}_{\overline{b_k}}|b_k\rangle}{\langle b_k|{(\hat\Pi^\pm)}_{\overline{b_k}}|b_k\rangle}
		-\frac{\langle b_k|{(\hat\Pi^\pm\hat L\hat\Pi^\pm)}_{\overline{b_k}}|b_k\rangle }{\langle b_k|{(\hat\Pi^\pm)}_{\overline{b_k}}|b_k\rangle^2}
		{(\hat\Pi^\pm)}_{\overline{b_k}}|b_k\rangle.
		\nonumber
	\end{align}
	Further, inserting the definition of $g$ in Eq.~\eqref{eq:RayleighQ} yields
	\begin{align}\label{eq:preSE}
		0=\frac{(\hat{\Pi}^{\pm}\hat{L}\hat{\Pi}^{\pm})_{\overline{b_j}}|b_j\rangle
		-g(\hat{\Pi}^\pm)_{\overline{b_j}}|b_j\rangle}{\langle b_j|(\hat{\Pi}^\pm)_{\overline{b_j}}|b_j \rangle}.
	\end{align}

	Let us summarize the conducted optimization which results in Eq.~\eqref{eq:preSE}.
	The derived set of algebraic equations,
	\begin{align}\label{eq:ISEequation_1stform}
		\big(\hat{\Pi}^{\pm}\hat{L}\hat{\Pi}^{\pm}\big)_{\overline{b_j}}|b_j\rangle = g\big(\hat{\Pi}^{\pm}\big)_{\overline{b_j}}|b_j\rangle
		\text{ for $j=1,\dots,K$,}
	\end{align}
	define the separability eigenvalue (SEvalue) equations for IP.
	The common eigenvalue $g$ is denoted as the SEvalue, and the (anti)symmetric product vector $\hat{\Pi}^{\pm} |b_1 , \dots , b_K\rangle$ is the corresponding separability eigenvector (SEvector) for IP.

	The SEvalue equations for IP represent a system of $K$ coupled eigenvalue equations.
	Remarkably, it turns out that they have the same structure as the corresponding equations for DP~\cite{MEW}.
	For DP, we use the $N$-fold identity $\hat 1$ that replaces the projector $\hat\Pi^\pm$ in~\eqref{eq:ISEequation_1stform}.
	Even though we find a strong relation to the case of DP, cf.~\cite{MEW,SSV14}, we want to point out that neither the distinguishable case includes the indistinguishable one, nor vice versa.
	This is due to the properties of the noninvertible operator $\hat\Pi^\pm$.

	We found that the SEvalue $g$ corresponds to an optimal expectation value of $\hat{L}$ for $K$-separable Boson or Fermion states.
	Therefore, we get the bound in the entanglement criterion~\eqref{eq:entanglement_condition_L} as
	\begin{align}
		\sup\{\mathrm{tr}( \hat{L} \hat{\sigma}):\text{for all $\hat{\sigma}$ in Eq.~\eqref{eq:separability_identical}}\}
		=\sup\{g\},
	\end{align}
	i.e., the initial convex optimization problem is solved by the largest SEvalue $g$ of the equations~\eqref{eq:ISEequation_1stform}.
	Now, the entanglement condition for the state $\hat\varrho$ may be written as
	\begin{align}
		\langle\hat L\rangle=\mathrm{tr}(\hat \varrho\hat L)>\sup\{g\}.
		\label{eq:entanglement_condition}
	\end{align}

	Alternatively, this condition can be written in terms of witnesses constructed from Eqs.~\eqref{eq:witness_representation_sym} and~\eqref{eq:witness_opt}:
	\begin{align}\label{eq:witness_g}
		\hat{W}_\pm = G \hat{\Pi}^{\pm} - \hat{\Pi}^{\pm}\hat{L}\hat{\Pi}^{\pm},
		\text{ with }G=\sup \left\lbrace g \right\rbrace,
	\end{align}
	which reads as $\mathrm{tr}(\hat \varrho\hat W_\pm)<0$.
	Similarly, a witness can be constructed using the lower bound of the Rayleigh quotient~\eqref{eq:RayleighQ} as
	\begin{align}
		\hat W_\pm=\hat\Pi^\pm\left[\hat L-\inf\{g\}\hat 1\right]\hat\Pi^\pm.
	\end{align}
	Hence, by solving the algebraic problem~\eqref{eq:ISEequation_1stform} of observables $\hat{L}$, we are able to construct, in principle, any optimal entanglement witnesses for multiple correlated Bosons or Fermions.
	Moreover, the SEvalue equations for IP might be also used for a numerical optimization if an analytical solution is not available.
	Since the criterion in~\eqref{eq:entanglement_condition} and the witnessing approach are equivalent, we study from now on solely the former one.

\subsection{Further properties of the SEvalue equations}

	Equivalent to the form of the SEvalue equations for IP in~\eqref{eq:ISEequation_1stform}, one might formulate a second form. 
	The solutions $g$ and $\hat{\Pi}^{\pm} |b_1 , \dots , b_K\rangle$ of the Hermitian operator $\hat{L}$ can be found by solving
	\begin{align}\label{eq:ISEeq_2ndform}
		\hat{\Pi}^{\pm}\hat{L} \hat{\Pi}^{\pm} |b_1 ,\dots , b_K\rangle  = g \hat{\Pi}^{\pm} |b_1 , \dots , b_K\rangle + |\chi\rangle,
	\end{align}
	with the perturbation term $|\chi\rangle$, which has to fulfill for all $j=1,\dots,K$ and for all $|x\rangle \in \mathcal{H}^{\otimes N_j}$ an orthogonality relation of the form
	\begin{align}
		\langle b_1,\dots,b_{j-1},x,b_{j+1},\dots,b_K | \chi\rangle = 0
	\end{align}
	to be equivalent with the first form in~\eqref{eq:ISEequation_1stform}.
	This treatment transforms the coupled set of equations of the first form~\eqref{eq:ISEequation_1stform} into a single, but perturbed, eigenvalue equation of the second form~\eqref{eq:ISEeq_2ndform}.
	Note that the perturbation $|\chi\rangle$ is an element of the (anti)symmetric subspace, since $\hat \Pi^\pm [\hat{L}\hat{\Pi}^{\pm}-g\hat 1]|b_1 ,\dots , b_K\rangle=|\chi\rangle$.

	Another important property of the SEvalue equations for IP is the behavior under certain transformations of the observable.
	Local unitaries $\hat{U}$ and shifts of $\hat L$, leading to a transformed observable
	\begin{align}
		\hat{L}' =\left[\hat{U}^{\otimes N}\right]^\dagger \left[ \lambda_1 \hat{L} + 
		\lambda_2 \hat{\Pi}^{\pm} \right] \left[\hat{U}^{\otimes N}\right],
	\end{align}
	with $\lambda_1$, $\lambda_2 \in \mathbb{R}  \backslash \lbrace 0 \rbrace$, can be directly passed onto the solutions.
	If the SEvalue $g$ together with the SEvector $\hat{\Pi}^{\pm} |b_1 , \dots , b_K\rangle$ is a solutions of the SEvalue equations for IP of $\hat L$, then the operator $\hat L'$ has the corresponding solutions:
	\begin{align}
		\text{SEvalue: }&g' = \lambda_1 g + \lambda_2,\\
		\nonumber\text{SEvector: }&
		\hat{\Pi}^{\pm} |b_1', \dots, b_K'\rangle = \hat{U}^{\otimes N} \hat{\Pi}^{\pm} |b_1 , \dots , b_K\rangle\\
		=& \hat{\Pi}^{\pm} \hat{U}^{\otimes N} |b_1 , \dots , b_K\rangle.
	\end{align}
	Hence, by solving the SEvalue equations for IP for a given $\hat L$ we gain the solutions for a whole class of observables.
	Additionally, we get for $\lambda_1=1$ and $\lambda_2=0$ that the SEvalues are invariant under local transformations.

\section{Bipartite example}\label{Sec:BipartiteExample}

	In a first application of our introduced method, we aim at witnessing bipartite entanglement.
	As our observable we may choose the rank one operator:
	\begin{align}\label{eq:rankone}
		\hat L=|\psi\rangle\langle\psi|,
	\end{align}
	being defined by a two-mode vector $|\psi\rangle\in\mathcal H\otimes\mathcal H$.
	For DP, we have the well-known Schmidt decomposition~\cite{QCQI} to represent this vector,
	\begin{align}\label{eq:rewritten-schmidtdecomp}
		|\psi\rangle=\sum_{i,j=1}^d\psi_{i,j}|i,j\rangle=\sum_{n=1}^d \lambda_n |u_n,v_n\rangle,
	\end{align}
	in terms of orthonormal sets $\{|u_n\rangle\}_{n=1}^d$ and $\{|v_n\rangle\}_{n=1}^d$ as well as non-negative coefficients $\lambda_n\geq0$.
	For a Fermion or Boson state we get the Slater decomposition -- as studied, for example, in Refs.~\cite{QCSIP,EMSIP,HJ13} -- as
	\begin{align}
		\nonumber |f\rangle =&\hat\Pi^{-}|\psi\rangle= \sum_{i,j=1}^d f_{i,j} |i,j\rangle\\
		=& \sum_{n=1}^{\lfloor d/2 \rfloor} \kappa_n (|w_{2n-1},w_{2n}\rangle-|w_{2n},w_{2n-1}\rangle),\label{eq:fermionic-schmidtdecomp}
		\\|b\rangle =&\hat\Pi^{+}|\psi\rangle= \sum_{i,j=1}^d b_{i,j} |i,j\rangle= \sum_{n=1}^{d} \kappa'_n |w'_{n},w'_{n}\rangle,\label{eq:bosonic-schmidtdecomp}
	\end{align}
	for $f_{i,j}=-f_{j,i}$, orthonormal $|w_n\rangle$, $\lfloor x \rfloor$ denoting the largest integer less or equal to $x$, and $\kappa_n\geq0$;
	and for $b_{i,j}=b_{j,i}$ with orthonormal $|w'_n\rangle$ and $\kappa'_n\geq0$.

	In Appendix~\ref{Sec:Example1}, the solution of the SEvalue equation for IP is explicitly computed for the considered observable~\eqref{eq:rankone}.
	Here, let us summarize the results.
	It is worth pointing out that the nontrivial solutions, i.e., $g\neq0$, have forms which are directly related to the decompositions~\eqref{eq:fermionic-schmidtdecomp} and~\eqref{eq:bosonic-schmidtdecomp}.
	Namely, we get for Fermions
	\begin{align}
		\text{SEvalues: }&g_n=2\kappa_n^2,\\
		\text{SEvectors: }&|w_{2n-1}\rangle\wedge|w_{2n}\rangle,
	\end{align}
	and for Bosons:
	\begin{align}
		\text{SEvalues: }&g_n={\kappa'_n}^2\text{ and }g_{k,l}={\kappa'_k}^2+{\kappa'_l}^2,\\
		\text{SEvectors: }&
		|w'_{n}\rangle\vee|w'_{n}\rangle
		\text{ and }
		|w^{+}_{k,l}\rangle\vee|w^{-}_{k,l}\rangle,
	\end{align}
	respectively, with $|w^{\pm}_{k,l}\rangle=\sqrt{\kappa'_k}|w'_k\rangle\pm i\sqrt{\kappa'_l}|w'_l\rangle$.
	Now, the entanglement condition~\eqref{eq:entanglement_condition}, can be written in terms of the fidelities:
	\begin{align}
		\langle b|\hat\varrho_{\rm B}|b\rangle>&\max_{1\leq k<l\leq d}\big\{{\kappa'_k}^2+{\kappa'_l}^2\big\}
		\\\text{or }
		\langle f|\hat\varrho_{\rm F}|f\rangle>&\max_{1\leq n\leq \lfloor d/2\rfloor}\big\{2\kappa_n^2\big\},
	\end{align}
	for bipartite, mixed or pure entangled states of Bosons, $\hat\varrho_{\rm B}$, or Fermions, $\hat\varrho_{\rm F}$.
	Note that, for the case of DP, we also get the separable bound from the decomposition~\eqref{eq:rewritten-schmidtdecomp} as $G=\max_{1\leq n \leq d}\{\lambda_n^2\}$, using the results in~\cite{SV09}:
	\begin{align}
		\text{SEvalues: }&g_n=\lambda_n^2,\\
		\text{SEvectors: }&|u_n\rangle\otimes|v_n\rangle.
	\end{align}

	Let us apply the method to characterize the entanglement of the pure state $|\psi\rangle$ which is mixed with white noise,
	\begin{align}\label{eq:noisy-pure-state}
		\hat{\rho} = p \hat{\mathbb{I}}|\psi\rangle\langle \psi|\hat{\mathbb{I}} + (1-p) \frac{\hat{\mathbb{I}}}{\text{tr } \hat{\mathbb{I}}},
	\end{align}
	with $p\in [0,1]$ being a noise parameter, $\hat{\mathbb I}\in\{\hat 1,\hat\Pi^{+},\hat\Pi^{-}\}$, and the second term being separable.
	For $p=0$, we get the separable state $\hat{\mathbb{I}}/\text{tr } \hat{\mathbb{I}}$, which is a uniformly weighted mixture of all normalized product state $\hat{\mathbb I}|a_1,a_2\rangle$ with $1=\langle a_1,a_2|\hat{\mathbb I}|a_1,a_2\rangle$; cf. Eqs.~\eqref{eq:separability} and~\eqref{eq:separability_identical}.
	Replacing $\hat{\mathbb I}$ with other projections we could additionally study entanglement for other parastatistics, e.g., for anyons~\cite{LM77,W82}, using the same treatment as presented for Bosons and Fermions.

	\begin{figure}[ht]
	\begin{center}
		\includegraphics[width=4.25cm]{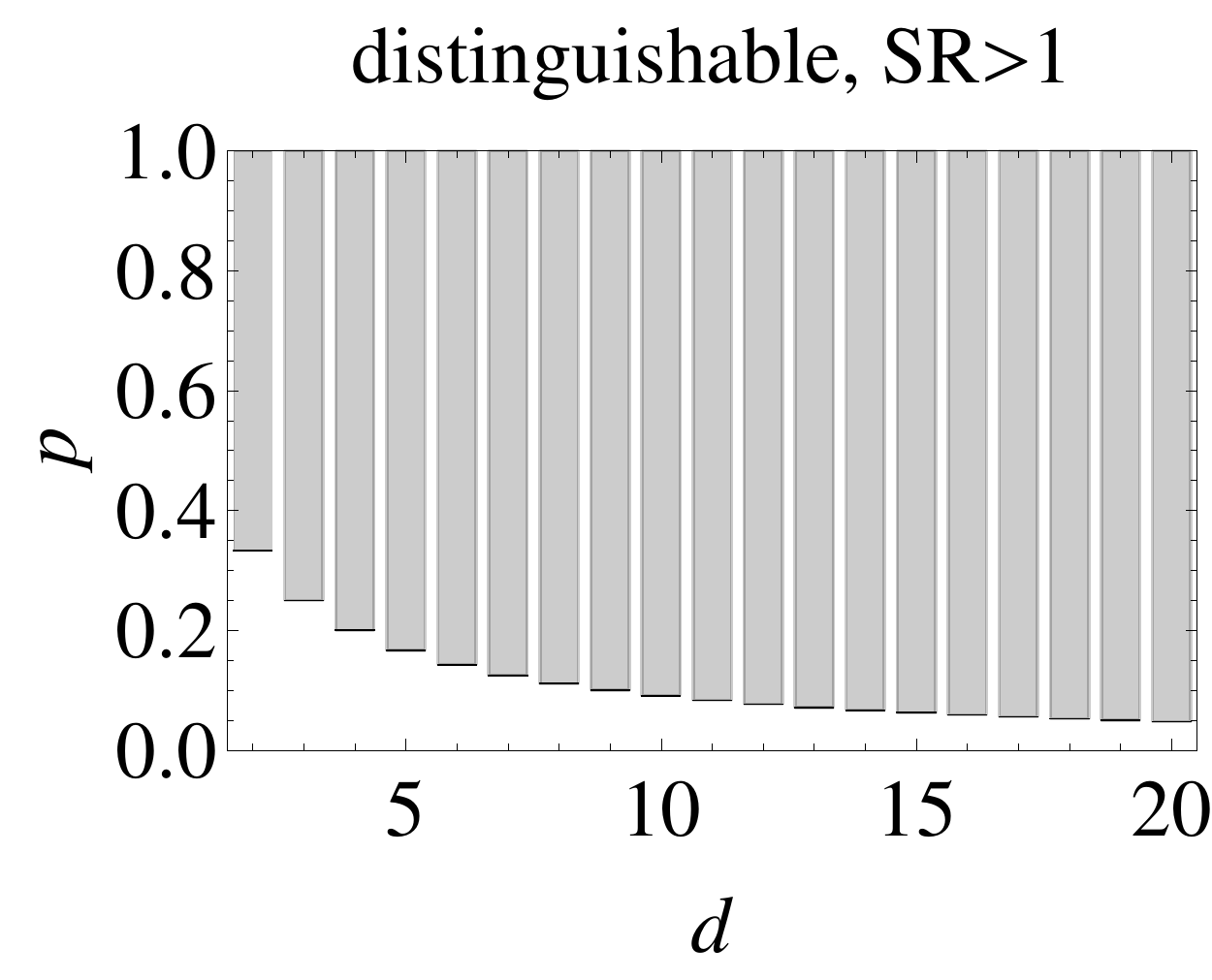}
		\includegraphics[width=4.25cm]{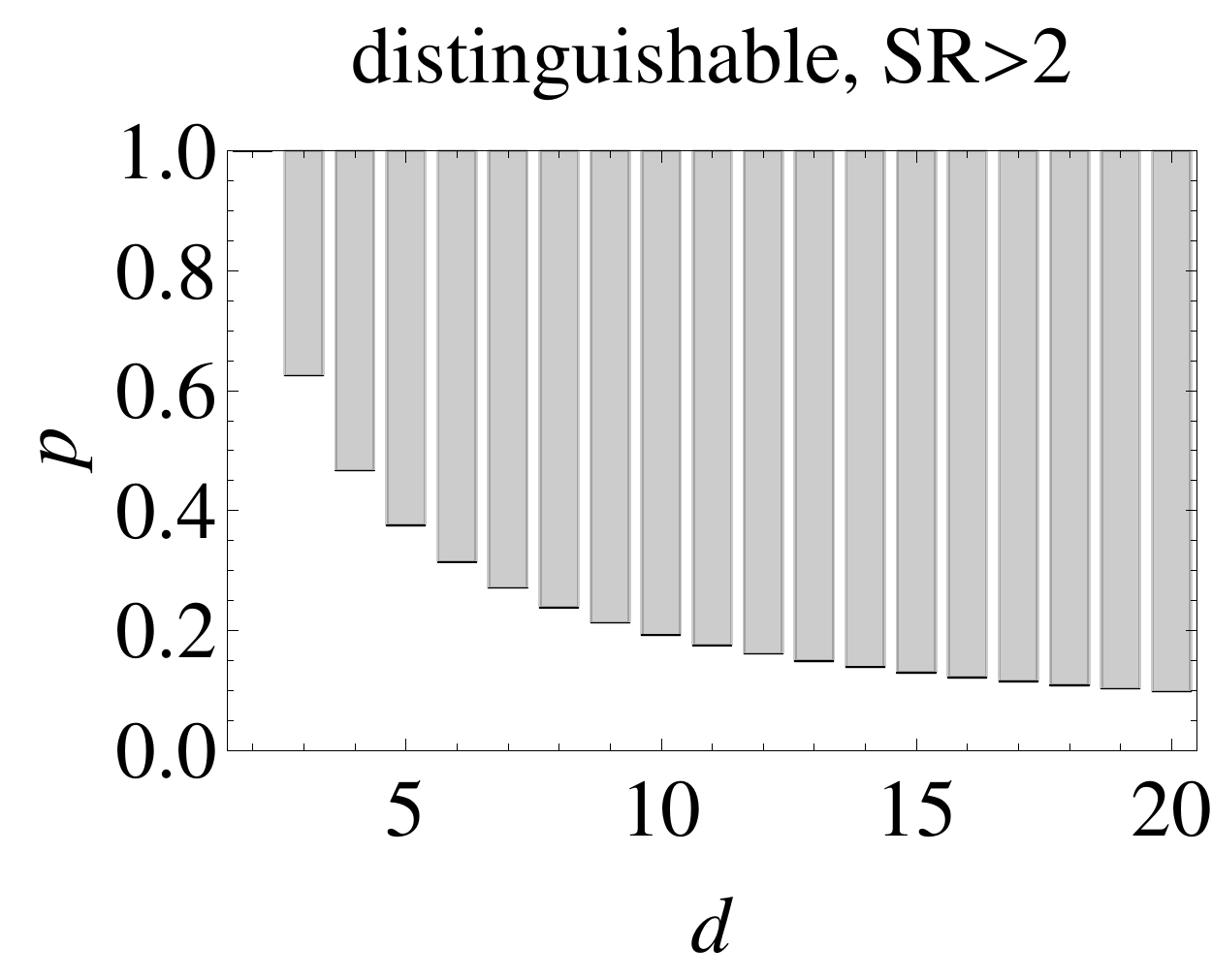}
		\\[0.2cm]
		\includegraphics[width=4.25cm]{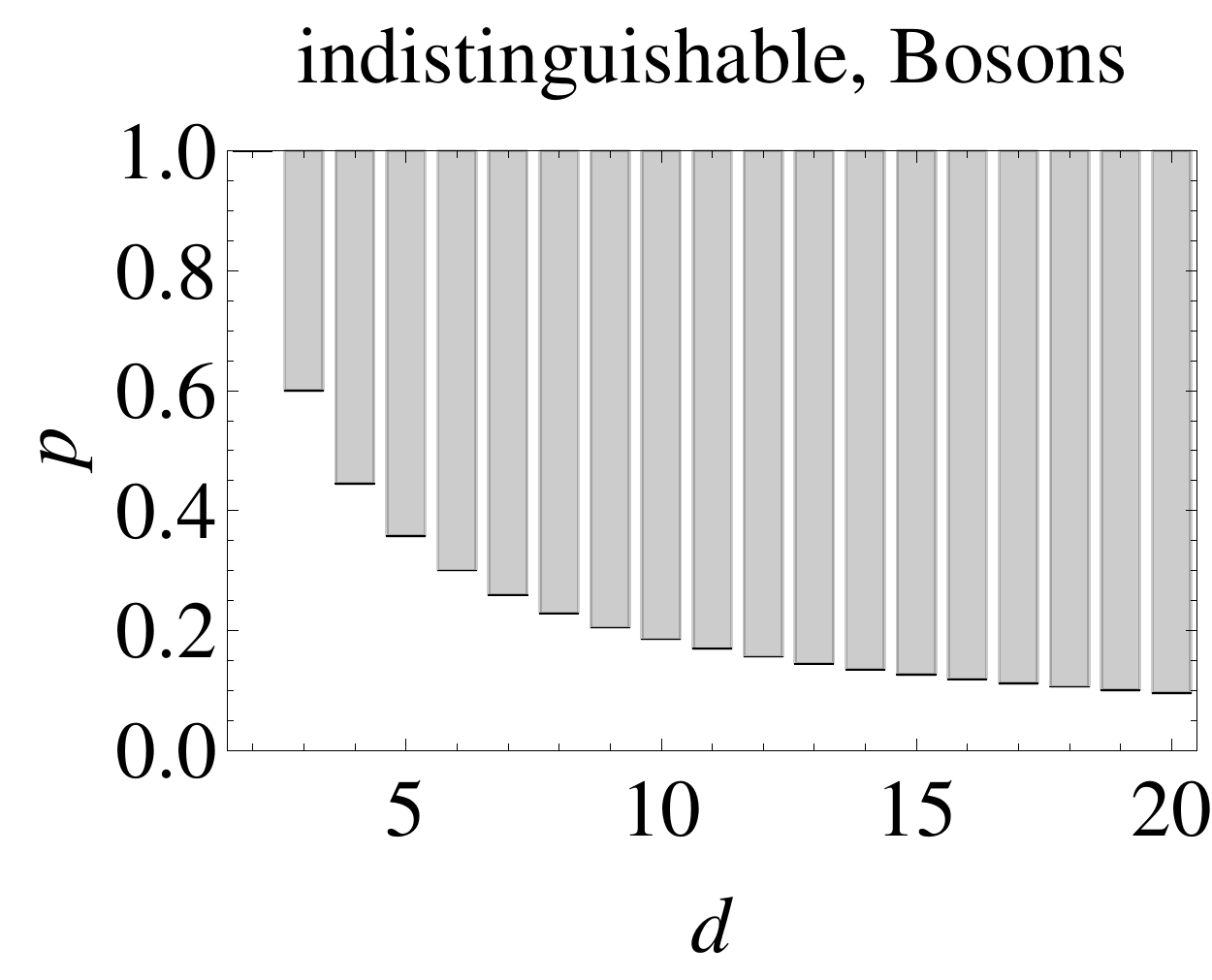}
		\includegraphics[width=4.25cm]{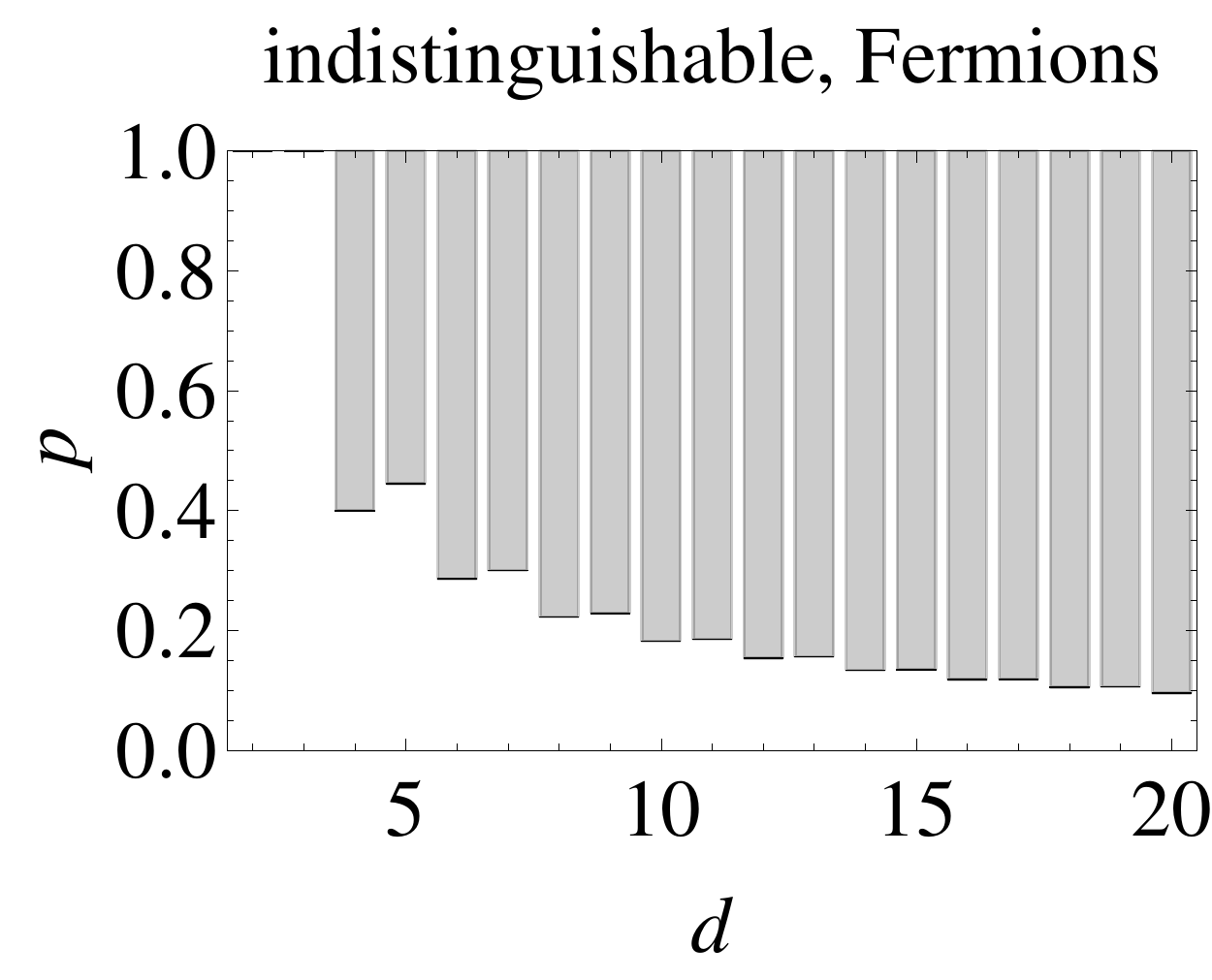}
		\caption{
			Mixing in terms of $p$ for the state~\eqref{eq:noisy-pure-state} depending on $d=\dim\mathcal H$ is plotted.
			As long as $p$ is in the gray shaded area, we successfully detected entanglement.
			The coefficients in Eqs.~\eqref{eq:rewritten-schmidtdecomp} and~\eqref{eq:bosonic-schmidtdecomp} for DP and Bosons, respectively, are chosen to be equal $\lambda_k=\kappa'_k=d^{-1/2}$ ($k=1,\ldots, d$).
			In the case of Fermions, we choose $\kappa_k=(2\lfloor d/2\rfloor)^{-1/2}$ ($k=1,\ldots, \lfloor d/2\rfloor$), see Eq.~\eqref{eq:fermionic-schmidtdecomp}, yielding a different behavior for even and odd dimensions $d$.
		}\label{fig:multiplot}
	\end{center}
	\end{figure}

	In Fig.~\ref{fig:multiplot}, we compare different quantum statistics regarding their entanglement properties for the mixed state~\eqref{eq:noisy-pure-state} in dependence on the dimensionality of the single particle's Hilbert space, $d=\dim\mathcal H$.
	We apply the test operator in~\eqref{eq:rankone}.
	As long as $\langle\hat L\rangle>\sup\{g\}$ (gray area in Fig.~\ref{fig:multiplot}), we have identified entanglement for the mixing parameter $p$ for DP (plot: SR${>}$1), Bosons or Fermions.
	Since the structure of (anti)symmetric product states is related to Bell-like states, cf. Eq.~\eqref{eq:KeyExample}, we also consider Schmidt rank (SR) two states for DP.
	The calculation of the corresponding bounds is done in~\cite{DSN} and applied in~\cite{DNLS}.
	For any $p$ in gray area of the plot SR${>}$2, we can conclude that more than two tensor-product states have to be superimposed to describe the state~\eqref{eq:noisy-pure-state}.
	Thus our approach allows the detection of different forms of entanglement based on a single observable.

\section{Multipartite example}\label{Sec:MultipartiteExample}
	In the following we will study a multipartite entanglement test, which is even independent of the spin statistics.
	We further assume $\dim\mathcal H=\infty$ given by the orthonormal single-mode basis $\{|n\rangle\}_{n=0}^\infty$.
	The observable is
	\begin{align}\label{eq:test-multi}
		\hat L{=}&|1,\dots,N\rangle\langle N{+}1,\dots,2N|
		{+}|N{+}1,\dots,2N\rangle\langle 1,\dots,N|.
	\end{align}
	For a state $\hat\rho$, the observable $\hat L$ measures an interference term of the form
	\begin{align}
		\nonumber \langle \hat L\rangle{=}&\langle N{+}1,\dots,2N|\hat\rho|1,\dots,N\rangle
		\\&{+}\langle 1,\dots,N|\hat\rho|N{+}1,\dots,2N\rangle.
	\end{align}
	In Appendix~\ref{Sec:Example2}, we solve the SEvalue equations for DP and IP.
	The obtained maximal bound for $K$-separable states is
	\begin{align}\label{eq:bound-multi}
		\sup\{g\}=\left(1/2\right)^{K-1}.
	\end{align}
	Note that this bound is even independent of the quantum statistics.
	Therefore, the entanglement condition~\eqref{eq:entanglement_condition_L} in this case states:
	Whenever the interference $\langle \hat L\rangle$ for $N$-particle system of DP, Bosons, or Fermions exceeds the bound $(1/2)^{K-1}$, we have certified that the state cannot be $K$ separable.

	The observable~\eqref{eq:test-multi} may be applied to a GHZ-type state~\cite{GHZ},
	\begin{align}\label{eq:GHZ-type}
		|q\rangle{=}\sqrt{\nu(\hat{\mathbb I})}\,\hat{\mathbb I}\sum_{n=0}^\infty \sqrt{1{-}|q|^2}q^n |nN{+}1,\dots,(n{+}1)N\rangle,
	\end{align}
	with $\hat{\mathbb I}\in\{\hat 1,\hat \Pi^{+},\hat \Pi^{-}\}$, $\nu(\hat\Pi^{\pm})=N!$, and $\nu(\hat 1)=1$.
	This state is of a GHZ-type structure, because for each mode $j$ holds that the individual vectors $|nN+j\rangle$ are orthonormal for different $n$.
	Using the transformations $\hat T_j|n\rangle{=}|nN{+}j\rangle$ in~\eqref{eq:trafo} of Appendix~\ref{Sec:Example2}, it can be directly seen that the state in~\eqref{eq:GHZ-type} is a GHZ-type of state,
	$|q\rangle=\hat{\mathbb I}(\hat T_1\otimes\dots\otimes\hat T_N)\sum_{n=0}^\infty \lambda_n |n,\dots,n\rangle$.
	In addition, the pure state might be perturbed due to a randomly distributed parameter $q$ ($|q|<1$):
	\begin{align}
		\hat\rho=\int_{|q|<1} d^2q\,p(q)|q\rangle\langle q|,
	\end{align}
	where $p$ is a classical probability distribution.

\begin{figure}[ht]
	\begin{center}
		\includegraphics[width=8.5cm]{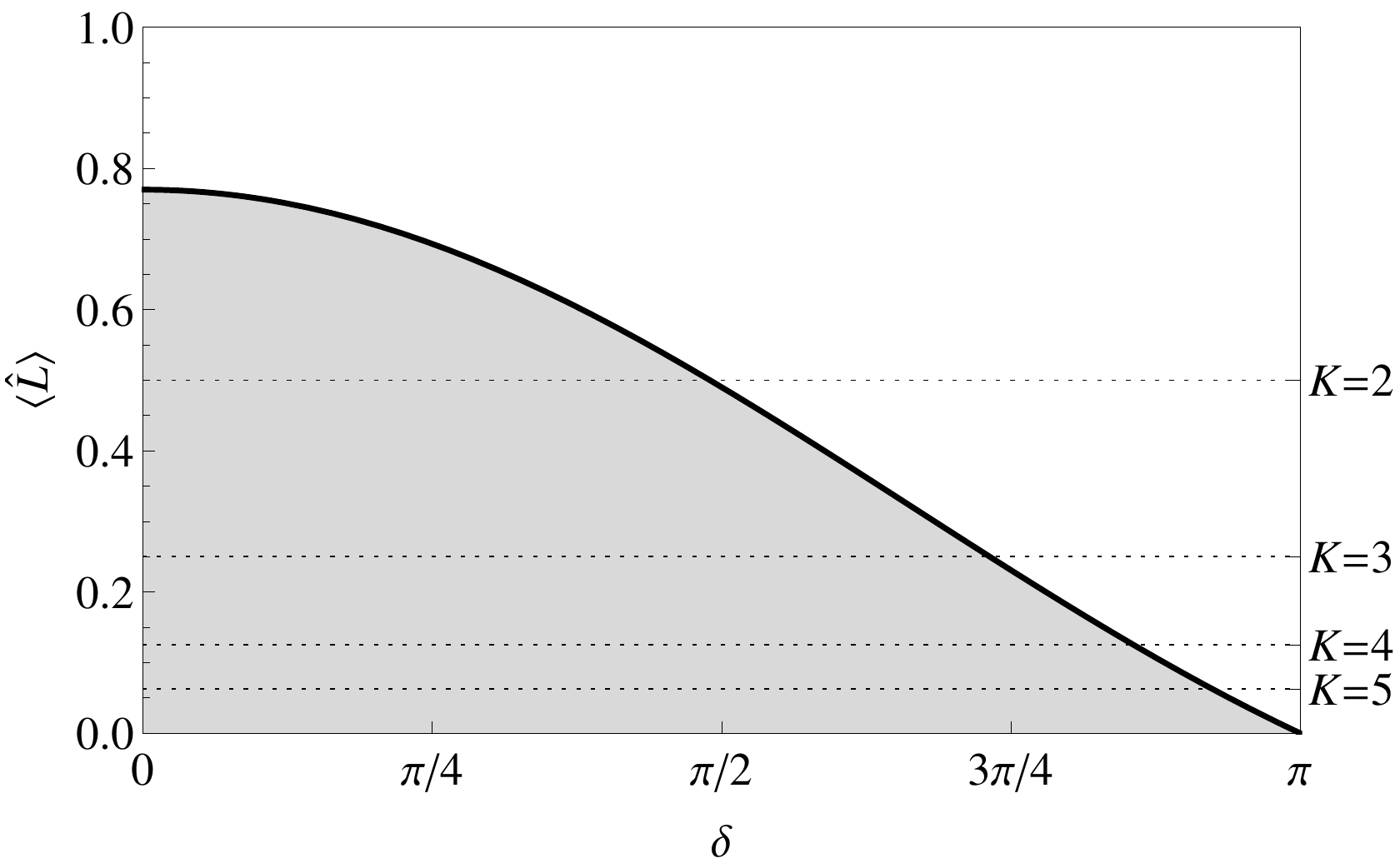}
		\caption{
			Expectation value in Eq.~\eqref{eq:expect-val-ksep} is plotted for $N=5$ (solid curve), an amplitude $|q|=1/\sqrt3$, and a uniformly distributed phase in the interval $\arg q\in[-\delta,\delta]$.
			If $\langle\hat L\rangle$ is above the dashed line $K$, then the state cannot be a $K$-separable one -- independent of the quantum statistics.
		}\label{fig:KsepPlot}
	\end{center}
	\end{figure}

	The identification of multipartite entangled Bosons and Fermions as well as DP is shown in Fig.~\ref{fig:KsepPlot} for a dephasing channel, i.e., the amplitude $r=|q|$ is fixed and phase $\varphi=\arg q$ is randomized.
	This uniform dephasing in the interval $\varphi\in [-\delta,+\delta]$ results in the density matrix
	\begin{align}
		\hat\rho=&\int_{-\delta}^{+\delta} \frac{d\varphi}{2\delta}\,|r\exp[{\rm i}\varphi]\rangle\langle r\exp[{\rm i}\varphi]|
		\\\nonumber =&\sum_{n,n'=0}^\infty (1{-}r^2) r^{n{+}n'}{\rm sinc}[\delta(n{-}n')]\nu(\hat{\mathbb I})
		\\\nonumber &\times \hat{\mathbb I}|nN{+}1,\dots,(n{+}1)N\rangle\langle n'N{+}1,\dots,(n'{+}1)N|\hat{\mathbb I},
	\end{align}
	with ${\rm sinc}[x]=\sin[x]/x$ (${\rm sinc}[0]=1$).
	Hence we have for this state, for all $\hat{\mathbb I}\in\{\hat 1,\hat\Pi^{+},\hat\Pi^{-}\}$, and for all $K$ partitions $(N_1,\dots,N_K)$ the entanglement condition
	\begin{align}\label{eq:expect-val-ksep}
		\langle\hat L\rangle=2(1{-}r^2) r\,{\rm sinc}[\delta]>(1/2)^{K-1}.
	\end{align}
	As long as the expectation value in Fig.~\ref{fig:KsepPlot} is above the dashed lines, we certified that the state cannot be a $K$-separable one.
	Note that for a full dephasing, $\delta=\pi$, this state is diagonal in product states and, therefore, separable.
	For no dephasing, $\delta=0$, we have a pure GHZ-type entangled state.
	This example demonstrates the general possibility to construct spin statistics independent entanglement tests with our approach.

\section{Conclusions}\label{Sec:Summary}
	In summary, we derived a method which allows the construction of entanglement probes in systems of Bosons and Fermions.
	These necessary and sufficient conditions are capable of determining full and partial entanglement for any number of particles.
	The optimization of these criteria is based on a set of generalized eigenvalue equations.
	These equations yield a structural unification of entanglement for distinguishable particles, Fermions and Bosons, and it can be generalized to other parastatistics.

	We analyzed and compared a number of examples for studying the differences and similarities between full and partial separability as well as entanglement in systems of distinguishable and indistinguishable particles.
	For instance, the determination of entanglement in bipartite and multipartite as well as discrete and continuous variable quantum systems demonstrate the wide range of applications of our technique even in the presence of noise.
	This also shows that our method is not limited to bipartitions or small numbers of particles.
	Moreover, the construction of spin-statistics independent entanglement probes has been established.
	Such witnesses can detect entanglement of a quantum system independent of symmetrization effects.

	We presented an approach which allows one to construct entanglement criteria, in principle, from almost all observables.
	However, the question which measurable quantity is able to witness the entanglement of a particular state is open and requires further studies.
	Beyond the here presented full analytical approach, numerical implementations may allow one to generate more sophisticated entanglement probes for detecting entanglement in more general scenarios.
	Therefore, we believe that our approach will provide a versatile tool to characterize entanglement in future experiments, with applications to Bose-Einstein condensates or ultra-cold Fermi systems.

\section*{Acknowledgments}
	This work was supported by the Deutsche Forschungsgemeinschaft through SFB~652.

\appendix

\section{Symmetrization of operators}\label{Sec:Subspaces}
	For an $N$-fold Hilbert space $\mathcal H^{\otimes N}$, the projection operators $\hat\Pi^{+}$ and $\hat\Pi^{-}$ are defined as
	\begin{align}
		\hat\Pi^\pm=\sum_{\sigma\in S_N} \frac{(\pm1)^{|\sigma|}}{N!}\hat P_\sigma,
	\end{align}
	with $\hat P_\sigma|a_1,\dots,a_N\rangle=|a_{\sigma(1)},\dots,a_{\sigma(N)}\rangle$ for any permutation $\sigma\in S_N$.
	It holds $(\hat\Pi^\pm)^\dagger=\hat\Pi^\pm$.
	We may study Hermitian operators in a product basis operator expansion, given by terms of the form
	\begin{align}
		\hat X=\hat Y_1\otimes\dots\otimes\hat Y_N, 
	\end{align}
	with $\hat Y_j=\hat Y_j^\dagger$ for $j=1,\dots,N$.
	The symmetric form of $\hat X$ is defined as
	\begin{align}
		\hat X^{\rm (sym)}=\sum_{\sigma\in S_N}\frac{1}{N!} \hat Y_{\sigma(1)}\otimes\dots\otimes\hat Y_{\sigma(N)}.
	\end{align}

	We claim
	\begin{align}\label{eq:symOP}
		\hat\Pi^\pm\hat X\hat\Pi^\pm=\hat X^{\rm (sym)}\hat\Pi^\pm=\hat\Pi^\pm\hat X^{\rm (sym)}.
	\end{align}
	The first equality can be directly computed, since for all $|a_1,\dots,a_N\rangle$ holds:
	\begin{align}
		\nonumber &\hat\Pi^\pm\hat X\hat\Pi^\pm\bigotimes_{j=1}^N |a_j\rangle=\hat\Pi^\pm\sum_{\sigma\in S_N}\frac{(\pm 1)^{|\sigma|}}{N!}\bigotimes_{j=1}^N \left[\hat Y_j|a_{\sigma(j)}\rangle\right]\\
		\nonumber =&\sum_{\sigma,\tau\in S_N}\frac{(\pm 1)^{|\sigma|+|\tau|}}{N!^2}\bigotimes_{j=1}^N \left[\hat Y_{\tau(j)}\right] \bigotimes_{j=1}^N|a_{\tau(\sigma(j))}\rangle\\
		\nonumber =&\sum_{\tau\in S_N}\frac{1}{N!} \bigotimes_{j=1}^N\hat Y_{\tau(j)}\sum_{\mu\in S_N}\frac{(\pm1)^{|\mu|}}{N!}\hat P_\mu\bigotimes_{j=1}^N |a_j\rangle\\
		=&\hat X^{\rm (sym)}\hat\Pi^\pm\bigotimes_{j=1}^N |a_j\rangle,
	\end{align}
	where we used a substitution $\mu=\tau\circ\sigma$ and $(\pm1)^{|\tau|+|\sigma|}=(\pm1)^{|\tau\circ\sigma|}$.
	The second equality in~\eqref{eq:symOP} follows from the fact that $\hat X$ and $\hat\Pi^\pm$ are Hermitian operators,
	\begin{align}
		\nonumber &\hat\Pi^\pm\hat X^{\rm (sym)}=\hat\Pi^\pm\hat X\hat\Pi^\pm=\big(\hat\Pi^\pm\hat X\hat\Pi^\pm\big)^\dagger\\
		=&\big(\hat\Pi^\pm\hat X^{\rm (sym)}\big)^\dagger=\hat X^{\rm (sym)}\hat\Pi^\pm.
	\end{align}

	For $\hat X=\hat 1\otimes\dots\otimes\hat 1$, we get from~\eqref{eq:symOP} that $\hat \Pi^\pm$ is idempotent.
	An observable $\hat L$ which solely acts on the corresponding subspaces ($\mathcal H^{\vee N}$ or $\mathcal H^{\wedge N}$) should fulfill the commutation relation $[\hat L,\hat \Pi^\pm]=0$.
	From~\eqref{eq:symOP} follows that this is fulfilled for every $\hat L=\hat L^{\rm (sym)}$.

\section{Existence of partial separable Bosons and Fermions}\label{Sec:PartSepIP}
	Let us prove that the set of partially separable states of IP includes more elements than the fully separable ones.
	For this reason, we consider the orthonormal basis $\{|0\rangle,\ldots,|4\rangle\}$ and the three-partite vector $|\Psi\rangle\in\left(\mathbb C^5\right){}^{\otimes 3}$ in Eq.~\eqref{Eq:AppPartSepDP}.
	It is partially separable in the tensor product, i.e., $|\Psi\rangle=|0\rangle\otimes|\Phi\rangle$.
	Applying the symmetrization or anti-symmetrization operator, $\hat\Pi^{\pm}$, we get
	\begin{align}\label{Eq:AppPartSepIP}
		|\Psi^{+}\rangle{=}\hat\Pi^{+}|\Psi\rangle{\in}\!\left(\mathbb C^5\right)^{\!\vee 3}
		\text{ and }
		|\Psi^{-}\rangle{=}\hat\Pi^{-}|\Psi\rangle{\in}\!\left(\mathbb C^5\right)^{\!\wedge 3},
	\end{align}
	or, more explicitly, we have the expansion
	\begin{align}
		|\Psi^{\pm}\rangle=&\frac{1}{6}[
		\nonumber |0,1,2\rangle\pm|0,2,1\rangle+|0,3,4\rangle\pm|0,4,3\rangle\\
		\nonumber &+|1,2,0\rangle\pm|1,0,2\rangle+|3,4,0\rangle\pm|3,0,4\rangle\\
		&+|2,0,1\rangle\pm|2,1,0\rangle+|4,0,3\rangle\pm|4,3,0\rangle
		]
		\\=&\frac{1}{6}[\nonumber
		|0\rangle\otimes|\Phi_0\rangle+
		|1\rangle\otimes|\Phi_1\rangle+
		|2\rangle\otimes|\Phi_2\rangle\\
		&+|3\rangle\otimes|\Phi_3\rangle+
		|4\rangle\otimes|\Phi_4\rangle],
	\end{align}
	with the orthogonal vectors
	$|\Phi_0\rangle=|1,2\rangle\pm|2,1\rangle+|3,4\rangle\pm|4,3\rangle$,
	$|\Phi_1\rangle=|2,0\rangle\pm|0,2\rangle$,
	$|\Phi_2\rangle=|0,1\rangle\pm|1,0\rangle$,
	$|\Phi_3\rangle=|4,0\rangle\pm|0,4\rangle$,
	and $|\Phi_4\rangle=|0,3\rangle\pm|3,0\rangle$.
	The reduced state -- tracing with respect to the first subsystem -- is
	\begin{align}
		&\hat \rho_{\rm red}={\rm tr}_1|\Psi^\pm\rangle\langle\Psi^\pm|
		\\\nonumber =&\frac{|\Phi_0\rangle\langle\Phi_0|{+}
		|\Phi_1\rangle\langle\Phi_1|{+}
		|\Phi_2\rangle\langle\Phi_2|{+}
		|\Phi_3\rangle\langle\Phi_3|{+}
		|\Phi_4\rangle\langle\Phi_4|}{6},
	\end{align}
	and it has a rank of five, ${\rm rank}(\hat\rho_{\rm red})=5$.

	For proving that the states $|\Psi^\pm\rangle$ cannot be fully separable, $|\Psi^+\rangle\ncong|a_1\rangle\vee|a_2\rangle\vee|a_3\rangle$ and $|\Psi^-\rangle\ncong|a_1\rangle\wedge|a_2\rangle\wedge|a_3\rangle$, let us study the properties of the reduced density matrix of fully separable Boson and Fermion states.
	We have
	\begin{align}
		\nonumber |s^\pm\rangle=&\hat\Pi^{\pm}(|a_1\rangle\otimes|a_2\rangle\otimes|a_3\rangle)\\
		=&\frac{1}{6}|a_1\rangle\otimes|s_1\rangle+
		\frac{1}{6}|a_2\rangle\otimes|s_2\rangle+
		\frac{1}{6}|a_3\rangle\otimes|s_3\rangle,
	\end{align}
	with $|s_1\rangle=|a_2,a_3\rangle\pm|a_3,a_2\rangle$, $|s_2\rangle=|a_3,a_1\rangle\pm|a_1,a_3\rangle$, and $|s_3\rangle=|a_1,a_2\rangle\pm|a_2,a_1\rangle$.
	Consequently, the range or image of the partially reduced operator is
	\begin{align}
		\mathcal R{=}{\rm Im}({\rm tr}_1|s^\pm\rangle\langle s^\pm|)\subseteq{\rm span}\{|s_1\rangle,|s_2\rangle,|s_3\rangle\}{=}\mathcal R',
	\end{align}
	where the linear span $\mathcal R'$ has a dimensionality of three (if $|s_1\rangle,|s_2\rangle,|s_3\rangle$ are linearly independent) or less.
	Since $\dim\mathcal R\leq 3$, it follows that the rank of the partially reduced operator of any fully separable state of IP is bounded by three:
	\begin{align}
		&{\rm rank}(\hat \rho'_{\rm red})\leq 3, \text{ for all}\\
		\nonumber &\hat \rho'_{\rm red}={\rm tr_1}\left[\hat\Pi^\pm|a_1,a_2,a_3\rangle\langle a_1,a_2,a_3|\hat\Pi^\pm\right].
	\end{align}
	Finally, we conclude that the states $|\Psi^\pm\rangle$ cannot be fully separable because of ${\rm rank}(\hat\rho_{\rm red})=5>3$.
	Thus, the states in~\eqref{Eq:AppPartSepIP}, using Eq.~\eqref{Eq:AppPartSepDP}, are authentic examples of partially separable states of Fermions and Bosons that are not fully separable.

\section{Bipartite observable}\label{Sec:Example1}
	In the following two subsections, the solution of the SEvalue equation for IP will be explicitly computed for the considered observable~\eqref{eq:rankone}.
	For simplicity, we will assume in the following $\hat L=|f\rangle\langle f|$ ($\hat L=|b\rangle\langle b|$) for the two-Fermion (two-Boson) system which yields $\hat L=\hat\Pi^{-}\hat L\hat\Pi^{-}$ ($\hat L=\hat\Pi^{+}\hat L\hat\Pi^{+}$).

\subsection{Solution for Fermions}
	First, we give a unitary state representation for arbitrary $d$-dimensional ($d\in\mathbb N\cup\{\infty\}$) pure states of two Fermions.
	We start with a Fermion state,
	\begin{align}
		|f\rangle = \sum_{i,j=1}^d f_{i,j} |i,j\rangle, \text{ with } f_{i,j} = - f_{j,i},
	\end{align}
	and introduce the skew-symmetric coefficient matrix
	\begin{align}
		\hat{M}_f = (f_{i,j})_{i,j}=-\hat{M}_f^{\rm T}.
	\end{align}
	Using the Autonne-Takagi factorization in Ref.~\cite{HJ13}, we find the Slater decomposition of this coefficient matrix,
	\begin{align}\label{eq:app:takagiantisym}
		&\hat{M}_f = \hat{U} \hat{D} \hat{U}^{\rm T},
		\text{ with } \hat{U}^{\dagger} \hat{U} = \hat{1}
		\\\nonumber &\text{ and ($d$ even): } \hat{D} = \bigoplus_{j=1}^{d/2} \kappa_j \begin{bmatrix}  0 & +1 \\ -1 & 0 \end{bmatrix},
		\\\nonumber &\text{ or ($d$ odd): } \hat{D} = \bigoplus_{j=1}^{(d-1)/2} \kappa_j \begin{bmatrix}  0 & +1 \\ -1 & 0 \end{bmatrix}\bigoplus [0],
	\end{align}
	with $\hat{D}$ being a block diagonal matrix containing anti-diagonal $2\times2$ blocks and $\kappa_j\geq0$.
	This yields Eq.~\eqref{eq:fermionic-schmidtdecomp} in the form
	\begin{align}\label{eq:FermiExpansion}
		|f\rangle=\hat U\otimes\hat U\sum_{n=1}^{\lfloor d/2\rfloor} \kappa_n (|2n-1,2n\rangle-|2n,2n-1\rangle),
	\end{align}
	with an orthonormal single-mode basis $\{|1\rangle,|2\rangle,\ldots\}$ and $\hat U|k\rangle=|w_k\rangle$.
	Since the SEvalues are invariant under unitary separable operations $\hat U\otimes\hat U$,  we assume, without loss of generality, that $\hat U=\hat 1$.

	Now, we consider more general operators having an expansion as
	\begin{align}
		\nonumber \hat L=\sum_{m,n=1}^{\lfloor d/2\rfloor} L_{m,n} &(|2m-1,2m\rangle-|2m,2m-1\rangle)\\
		&\times(\langle 2n-1,2n|-\langle 2n,2n-1|),
	\end{align}
	which includes the special case $L_{m,n}=\kappa_m\kappa_n$ of projection operator $\hat L=|f\rangle\langle f|$.
	Since $\hat L=\hat\Pi^{-}\hat L\hat\Pi^{-}$, the SEvalue equations for Fermions in the second form~\eqref{eq:ISEeq_2ndform} read as
	\begin{align}
		\hat L|a_1,a_2\rangle=g\frac{1}{2}(|a_1,a_2\rangle-|a_2,a_1\rangle)+|\chi\rangle.
	\end{align}
	Using $\gamma_n=(\langle 2n-1,2n|-\langle 2n,2n-1|)|a_1,a_2\rangle$, we get
	\begin{align}
		&\hat L|a_1,a_2\rangle\\
		\nonumber =&\sum_{m=1}^{\lfloor d/2\rfloor}\left[\sum_{n=1}^{\lfloor d/2\rfloor} L_{m,n}\gamma_n\right](|2m-1,2m\rangle-|2m,2m-1\rangle).
	\end{align}
	We find that $\hat L|a_1,a_2\rangle$ is already diagonalized in the form~\eqref{eq:FermiExpansion}.
	Hence, the orthogonality of $|a_1,a_2\rangle$ to the perturbation $|\chi\rangle$ is fulfilled if
	\begin{align}
		\nonumber \hat\Pi^{-}|a_1,a_2\rangle=&\frac{1}{2}(|2n{-}1,2n\rangle{-}|2n,2n{-}1\rangle)
		{\cong}|2n{-}1\rangle{\wedge}|2n\rangle,
		\\g=&2L_{n,n},
		\\\nonumber |\chi\rangle=&\sum_{m\neq n}L_{m,n}(|2m-1,2m\rangle-|2m,2m-1\rangle),
	\end{align}
	for all $n = 1,\dots, \lfloor d/2 \rfloor$.
	For the special case $L_{m,n}=\kappa_m\kappa_n$, we get the maximal SEvalue as
	\begin{align}
		G=\max_{n}\{2 \kappa_n^2\}.
	\end{align}
	Note that trivial solutions, $g=0$, can be obtained by $|k\rangle\wedge|l\rangle$ with $(k,l)\neq(2n,2n+1)$.

\subsection{Solution for Bosons}
	Again, we first give the state representation according to Ref.~\cite{HJ13} for arbitrary $d$-dimensional pure state of two Bosons.
	This means that the symmetric state
	\begin{align}
		|b\rangle = \sum_{i,j=1}^d \, b_{i,j} |i,j\rangle, \text{ with } b_{i,j} = b_{j,i},
	\end{align}
	can be identified with a symmetric coefficient matrix
	\begin{align}\label{eq:app:takagi_sym}
		\hat{M}_b = (b_{i,j})_{i,j}=\hat{M}_b^{\rm T}
		\text{ and }
		\hat{M}_b = \hat{U} \hat{D} \hat{U}^{\rm T},
	\end{align}
	with $\hat{U}^{\dagger} \hat{U} = \hat{1}$ and $\hat{D} = {\rm diag} [\kappa'_1, \dots , \kappa'_d]\geq0$.
	Thus the symmetric Slater representation~\eqref{eq:bosonic-schmidtdecomp} of the state is
	\begin{align}\label{eq:symmetric_decomp}
		|b\rangle = \hat{U}\otimes \hat{U}\sum_{n=1}^d \kappa'_n |n,n\rangle,
		\text{ with } \hat U|n\rangle=|w'_n\rangle.
	\end{align}

	As in the previous example for Fermions, let us consider the more general operator
	\begin{align}
		\hat L=\sum_{m,n=1}^d L_{m,n}|m,m\rangle\langle n,n|.
	\end{align}
	The relation $\hat L=\hat\Pi^{+}\hat L\hat\Pi^{+}$ simplifies the SEvalue equations for Bosons in the second form to
	\begin{align}\label{eq:BosonN2SE}
		\hat L|a_1,a_2\rangle=g\frac{1}{2}(|a_1,a_2\rangle+|a_2,a_1\rangle)+|\chi\rangle.
	\end{align}
	Using $\gamma_n=\langle n,n|a_1,a_2\rangle$, we can now write
	\begin{align}\label{eq:TempSolBoson}
		\hat L|a_1,a_2\rangle
		=\sum_{m=1}^{d}\left[\sum_{n=1}^{d} L_{m,n}\gamma_n\right]|m,m\rangle.
	\end{align}
	Hence one class of solutions with $|a_1\rangle=|a_2\rangle$ is given by
	\begin{align}
		\nonumber \hat\Pi^{+}|a_1,a_2\rangle=&|n,n\rangle\cong|n\rangle\vee|n\rangle,
		\\g=&L_{n,n},
		\\\nonumber |\chi\rangle=&\sum_{m\neq n}L_{m,n}|m,m\rangle.
	\end{align}

	Unlike in the Fermion case, we have to take a brief look on the decomposition of product states of Bosons.
	Namely the state $\hat \Pi^+|a_1,a_2\rangle$ for any $|a_1\rangle\neq|a_2\rangle$ has a decomposition, cf. Eq.~\eqref{eq:symmetric_decomp}, as
	\begin{align}\label{eq:ProductBosonDecomp}
		\hat \Pi^+|a_1,a_2\rangle=&\hat U'\otimes \hat U'(\lambda'_1|1,1\rangle+\lambda_2'|2,2\rangle),
		\\\text{for }|a_{1(2)}\rangle=&\hat U'\left[\sqrt{\lambda'_1}|1\rangle{+}({-}){\rm i}\sqrt{\lambda'_2}|2\rangle\right]\nonumber.
	\end{align}
	Hence, we get a more involved set of solutions of Eq.~\eqref{eq:TempSolBoson} in the form (for $k\neq l$):
	\begin{align}
		\hat\Pi^{+}|a_1,a_2\rangle=&\lambda'_k|k,k\rangle+\lambda_l'|l,l\rangle, \label{eq:BoseSolvePre1}
		\\|\chi\rangle=&\sum_{m\neq k,l}(L_{m,k}\lambda'_k+L_{m,l}\lambda'_l)|m,m\rangle, \label{eq:BoseSolvePre2}
	\end{align}
	where the coefficients $\lambda'_k$ and $\lambda'_l$ have to be determined.
	We insert~\eqref{eq:BoseSolvePre1} and~\eqref{eq:BoseSolvePre2} into~\eqref{eq:BosonN2SE},
	\begin{align}
		\hat L\hat\Pi^{+}|a_1,a_2\rangle-|\chi\rangle=g\hat\Pi^{+}|a_1,a_2\rangle,
	\end{align}
	and find that the remaining terms to be computed are
	\begin{align}
		(L_{k,k}\lambda'_k+L_{k,l}\lambda'_l)|k,k\rangle&\nonumber\\
		+(L_{l,k}\lambda'_k+L_{l,l}\lambda'_l)|l,l\rangle&=g(\lambda'_k|k,k\rangle+\lambda_l'|l,l\rangle).
	\end{align}
	This is a standard eigenvalue problem in $\mathbb C^2$, which has the solutions
	\begin{align}
		g^{\pm} =& \frac{ L_{k,k} + L_{l,l} \pm \Delta }{2},\,\nonumber
		\Delta=\sqrt{ (L_{k,k} - L_{l,l})^2+4|L_{k,l}|^2 },
		\\\lambda'_k=&2L_{k,l}
		\text{ and }\lambda'_l=L_{l,l}-L_{k,k}\pm\Delta,
	\end{align}
	with the Hermiticity condition $L_{l,k}=L_{k,l}^\ast$.

	Again, in the particular case $L_{m,n}=\kappa'_m\kappa'_n$, we get the simplified solutions
	\begin{align}
		g=\lambda_n^2,\, g^{-}=0, \text{ and } g^{+}={\kappa'_k}^2+{\kappa'_l}^2.
	\end{align}
	Combining the solutions of the form $|a_1\rangle\vee|a_1\rangle$ and $|a_1\rangle\vee|a_2\rangle$ as well as using the fact that ${\kappa'_k}^2+{\kappa'_l}^2\geq {\kappa'_l}^2$, we get the maximal SEvalue as
	\begin{align}
		G=\max_{k\neq l}\{\lambda_k^2+\lambda_l^2\}.
	\end{align}

\section{Multipartite observable}\label{Sec:Example2}
	We considered an interference operator $\hat L$ in Eq.~\eqref{eq:test-multi} whose expectation value is the real part of an off-diagonal element of the density operator $\hat\rho$, $\langle \hat L\rangle=2\,{\rm Re}(\rho_{(1,\dots,N),(N{+}1,\dots,2N)})$. 
	Local unitary operations allow the generalization to other off-diagonal elements or phase shifts, ${\rm Re}(\exp[{\rm i}\varphi]\rho_{(1,\dots,N),(N{+}1,\dots,2N)})$.

	The injective transformations $\hat T_j$ of the orthonormal basis $\{|n\rangle\}_{n\in\mathbb N}$,
	\begin{align}\label{eq:trafo}
		\hat T_j|n\rangle{=}|nN{+}j\rangle \text{ for }j=1,\dots,N,
	\end{align}
	are constructed such that one can directly see that the for all $j,j'=1,\dots,N$ and $n,n'\in\mathbb N$ an orthogonality is given, $\langle n N+j| n' N+j'\rangle=\delta_{n,n'}\delta_{j,j'}$.
	Therefore, we get for $\hat{\mathbb I}\in\{\hat 1,\hat\Pi^{+},\hat\Pi^{-}\}$ the orthogonality relation
	\begin{align}\label{eq:special-orthonormality}
		\langle 1,\dots,N|\hat{\mathbb I}|N{+}1,\dots,2N\rangle=0
	\end{align}
	as well as the normalizations
	\begin{align}
		&\langle N{+}1,\dots,2N|\hat{\mathbb I}|N{+}1,\dots,2N\rangle \nonumber\\
		=&\langle 1,\dots,N|\hat{\mathbb I}|1,\dots,N\rangle=1/\nu(\hat{\mathbb I}),
	\end{align}
	with $\nu(\hat\Pi^{\pm})=N!$ and $\nu(\hat 1)=1$.
	Due to this fact, we may define the $K$-separable vectors
	\begin{align}
	\begin{aligned}
		|v_1,\dots,v_K\rangle{=}&|1,\dots,N\rangle
		\\\text{and }
		|w_1,\dots,w_K\rangle{=}&|N{+}1,\dots,2N\rangle,
	\end{aligned}
	\end{align}
	which are orthogonal for Fermions, Bosons, and DP and any partition $(N_1,\dots,N_K)$.

	As a last fact before we solve the SEvalue equations for this operator, let us recall an example of the standard eigenvalue problem:
	\begin{align}\label{eq:std-eigenvalue-example}
		\hat M=&m|m_w\rangle\langle m_v|+m^\ast|m_v\rangle\langle m_w|,\\
		\label{eq:std-eigenvalue-example1}|\mu_\pm\rangle=&\left(|m_w\rangle\pm\frac{m^\ast}{|m|}|m_v\rangle\right)/\sqrt 2,\\
		\label{eq:std-eigenvalue-example2}\mu_\pm=&\pm|m|,
	\end{align}
	with complex $m\neq 0$, orthonormal $\{|m_w\rangle,|m_v\rangle\}$, and $|\mu_\pm\rangle$ being the eigenvectors of $\hat M$ to the eigenvalues $\mu_\pm$.

	Now, let us use the first form of the SEvalue equation for IP and DP of the operator~\eqref{eq:test-multi}.
	Since the spanned subspace of $\hat L$ is ${\rm span}\{|v_1,\dots,v_K\rangle,|w_1,\dots,w_K\rangle\}$, let us expand
	\begin{align}
		|b_j\rangle=\beta_{v,j}|v_j\rangle+\beta_{w,j}|w_j\rangle.
	\end{align}
	We get for the $j$th SEvalue equation the two components
	\begin{align}
	\begin{aligned}
		\langle v_j|\big(\hat{\mathbb I}\hat L\hat{\mathbb I}\big)_{\overline{b_j}}|b_j\rangle=&g\langle v_j|\big(\hat{\mathbb I}\big)_{\overline{b_j}}|b_j\rangle,
		\\\langle w_j|\big(\hat{\mathbb I}\hat L\hat{\mathbb I}\big)_{\overline{b_j}}|b_j\rangle=&g\langle w_j|\big(\hat{\mathbb I}\big)_{\overline{b_j}}|b_j\rangle.
	\end{aligned}
	\end{align}
	Equivalently, we get by a rescaling with $\nu(\hat{\mathbb I})$
	\begin{align}
	\begin{aligned}
		\prod_{i\neq j}\left(\beta_{v,i}^\ast\beta_{w,i}\right)\,\beta_{w,j}{=}&g\,\prod_{i\neq j}\left(|\beta_{v,i}|^2+|\beta_{w,i}|^2\right)\,\beta_{v,j},\\
		\prod_{i\neq j}\left(\beta_{w,i}^\ast\beta_{v,i}\right)\,\beta_{v,j}{=}&g\,\prod_{i\neq j}\left(|\beta_{v,i}|^2+|\beta_{w,i}|^2\right)\,\beta_{w,j},
	\end{aligned}
	\end{align}
	which has the structure of the eigenvalue problem in~\eqref{eq:std-eigenvalue-example} with the solution in Eqs.~\eqref{eq:std-eigenvalue-example1} and~\eqref{eq:std-eigenvalue-example2}.
	Hence for each $j$ we get the solution for components with $|\beta_{v,j}|=|\beta_{w,j}|=1/\sqrt{2}$, yielding $|\beta_{v,i}|^2+|\beta_{w,i}|^2=1$ and the eigenvalues
	\begin{align}
		g=\pm\Big|\prod_{i\neq j}\left(\beta_{w,i}^\ast\beta_{v,i}\right)\Big|=\pm\left(1/2\right)^{K-1}.
	\end{align}
	Note for $\prod_{i\neq j}\left(\beta_{v,i}^\ast\beta_{w,i}\right)=0$, we get the trivial SEvalue $g=0$ and, for example, the SEvector $\hat{\mathbb I}|b_1,\dots,b_K\rangle=\hat{\mathbb I}|v_1,\dots,v_K\rangle$.
	Finally, the maximal SEvalue is
	\begin{align}
		G=\sup\{g\}=\left(1/2\right)^{K-1}.
	\end{align}
	Note that this result is independent of the particular $K$-partition $(N_1,\dots,N_K)$ and, due to especially chosen orthonormality in~\eqref{eq:special-orthonormality}, the result is also independent of the spin statistics.



\begin{thebibliography}{99}
\bibitem{EPR}  		A. Einstein, B. Podolsky, and N. Rosen,
			{\it Can Quantum-Mechanical Description of Physical Reality Be Considered Complete?},
			Phys. Rev. \textbf{47}, 777 (1935).
\bibitem{SCHROE1}	E. Schr\"odinger,
			{\it Discussion of probability relations between separated systems},
			Proc. Cambr. Philos. Soc. \textbf{31}, 555 (1935).
\bibitem{SCHROE2}	E. Schr\"odinger,
			{\it Probability relations between separated systems},
			Proc. Cambr. Philos. Soc. \textbf{32}, 446 (1936).
\bibitem{Review1}	N. Brunner, D. Cavalcanti, S. Pironio, V. Scarani, and S. Wehner,
			{\it Bell nonlocality},
			Rev. Mod. Phys. {\bf 86}, 419 (2014).
\bibitem{QM2}		V. Giovannetti, S. Lloyd, and L. Maccone,
			{\it Quantum-Enhanced Measurements: Beating the Standard Quantum Limit},
			Science \textbf{306}, 1330 (2004).
\bibitem{QM3}		P. M. Anisimov, G. M. Raterman, A. Chiruvelli, W. N. Plick, S. D. Huver, H. Lee, and J. P. Dowling,  
			{\it Quantum Metrology with Two-Mode Squeezed Vacuum: Parity Detection Beats the Heisenberg Limit},
			Phys. Rev. Lett. \textbf{104}, 103602, (2010).
\bibitem{QM1}  		V. Giovannetti, S. Lloyd, and L. Maccone,
			{\it Advances in quantum metrology},
			Nature Photon. \textbf{5}, 222 (2011).
\bibitem{QM4}  		B. M. Escher, R. L. de Matos Filho, and L. Davidovich,
			{\it General framework for estimating the ultimate precision limit in noisy quantum-enhanced metrology},
			Nat. Phys. \textbf{7}, 406 (2011).
\bibitem{QCQI} 		M. A. Nielsen and I. L. Chuang, 
			{\it Quantum Computation and Quantum Information},
			(Cambridge University Press, Cambridge, UK, 2000).
\bibitem{QE1}		R. Horodecki, P. Horodecki, M. Horodecki, and K. Horodecki,
			{\it Quantum entanglement},
			Rev. Mod. Phys. {\bf 81}, 865 (2009).
\bibitem{QP1}		O. G\"uhne and G. T\'{o}th,
			{\it Entanglement detection},
			Phys. Rep. \textbf{474}, 1 (2009).
\bibitem{SPAIEF}	A. M. L. Messiah and O. W. Greenberg,
			{\it Symmetrization Postulate and Its Experimental Foundation},
			Phys. Rev. \textbf{136}, B248 (1964).
\bibitem{EIP}		F. Benatti, R. Floreanini, and K. Titimbo,
			{\it Entanglement of Identical Particles},
			Open Syst. Inf. Dyn. {\bf 21}, 1440003 (2014).
\bibitem{RFW}		R. F. Werner,
			{\it Quantum states with EPR correlations admitting a hidden-variable model},
			Phys. Rev. A \textbf{40}, 4277 (1989).
\bibitem{QCTFS} 	J. Schliemann, J. Ignacio Cirac, M. Ku\'{s}, M. Lewenstein, and D. Loss,
			{\it Quantum Correlations in Two-Fermion Systems},
			Phys. Rev. A \textbf{64}, 022303 (2001).
\bibitem{GMW02}		G. Ghirardi, L. Marinatto, and T. Weber,
			{\it Entanglement and Properties of Composite Quantum Systems: a Conceptual and Mathematical Analysis},
			J. Stat. Phys. {\bf 108}, 49 (2002).
\bibitem{GM04}		G. Ghirardi and L. Marinatto,
			{\it General criterion for the entanglement of two indistinguishable particles},
			Phys. Rev. A {\bf 70}, 012109 (2004).
\bibitem{PY01}		R. Pa\v{s}kauskas and L. You,
			{\it Quantum correlations in two-boson wave functions},
			Phys. Rev. A {\bf 64}, 042310 (2001).
\bibitem{QCSIP}  	K. Eckert, J. Schliemann, D. Bru\ss{}, and M. Lewenstein,
			{\it Quantum Correlations in Systems of Indistinguishable Particles},
			Ann. Phys. (N.Y.) \textbf{299}, 88 (2002).
\bibitem{SMS}		M. Horodecki, P. Horodecki, and R. Horodecki, 
			{\it Separability of Mixed States: Necessary and Sufficient Conditions},
			Phys. Lett. A \textbf{223}, 1 (1996).
\bibitem{SMSn}		M. Horodecki, P. Horodecki, and R. Horodecki,
			{\it Separability of n-particle mixed states: necessary and sufficient conditions in terms of linear maps},
			Phys. Lett. A {\bf 283}, 1 (2001).
\bibitem{BWKGWGHBLS04}	M. Bourennane, M. Eibl, C. Kurtsiefer, S. Gaertner, H. Weinfurter, O. G\"uhne, P. Hyllus, D. Bru\ss{}, M. Lewenstein, and A. Sanpera,
			{\it Experimental Detection of Multipartite Entanglement using Witness Operators},
			Phys. Rev. Lett. {\bf 92}, 087902 (2004).
\bibitem{VCMM09} 	G. Vallone, R. Ceccarelli, F. De Martini, and P. Mataloni,
			{\it Hyperentanglement of two photons in three degrees of freedom},
			Phys. Rev. A {\bf 79}, 030301(R) (2009).
\bibitem{DRSJCGFGDS10}	L. DiCarlo, M. D. Reed, L. Sun, B. R. Johnson, J. M. Chow, J. M. Gambetta, L. Frunzio, S. M. Girvin, M. H. Devoret, and R. J. Schoelkopf,
			{\it Preparation and Measurement of Three-Qubit Entanglement in a Superconducting Circuit},
			Nature (London) {\bf 467}, 574 (2010).
\bibitem{JNKGLGCCP10}	B. Jungnitsch, S. Niekamp, M. Kleinmann, O. G\"uhne, H. Lu, W.-B. Gao, Y.-A. Chen, Z.-B. Chen, and J.-W. Pan,
			{\it Increasing the Statistical Significance of Entanglement Detection in Experiments},
			Phys. Rev. Lett. {\bf 104}, 210401 (2010).
\bibitem{AGDLRL13}	J. M. Arrazola, O. Gittsovich, J. M. Donohue, J. Lavoie, K. J. Resch, and N. L\"utkenhaus,
			{\it Reliable entanglement verification},
			Phys. Rev. A {\bf 87}, 062331 (2013).
\bibitem{DLTEK14}	J. Dai, Y. L. Len, Y. S. Teo, B.-G. Englert, and L. A. Krivitsky,
			{\it Experimental Detection of Entanglement with Optimal-Witness Families},
			Phys. Rev. Lett. {\bf 113}, 170402 (2014).
\bibitem{SSN}		F. Benattia, R. Floreaninib, and U. Marzolinoa,
			{\it Sub-shot-noise quantum metrology with entangled identical particles},
			Ann. Phys. (N.Y.) \textbf{325}, 924 (2010).
\bibitem{EoIP}		T. Sasaki, T. Ichikawa, and I. Tsutsui,
			{\it Entanglement of Indistinguishable Particles},
			Phys. Rev. A \textbf{83}, 012113 (2011).
\bibitem{MTE}		F. Buscemi and P. Bordone,
			{\it A measure of tripartite entanglement in bosonic and fermionic systems},
			Phys. Rev. A \textbf{84}, 022303 (2011).
\bibitem{ERSIP}		F. Benattia, R. Floreaninib, and U. Marzolinoa,
			{\it Entanglement robustness and geometry in systems of identical particles},
			Phys. Rev. A \textbf{85}, 042329 (2012).
\bibitem{BESIP}		F. Benattia, R. Floreaninib, and U. Marzolinoa,
			{\it Bipartite entanglement in systems of identical particles: the partial transposition criterion},
			Ann. Phys. (N.Y.) \textbf{327}, 1304 (2012).
\bibitem{OK13}		M. Oszmaniec and M. Ku\'s,
			{\it Universal framework for entanglement detection},
			Phys. Rev. A {\bf 88}, 052328 (2013);
			M. Oszmaniec and M. Ku\'s,
			{\it Fraction of isospectral states exhibiting quantum correlations},
			arXiv:1312.7359 [quant-ph].
\bibitem{USEIP}		T. Sasaki, T. Ichikawa, and I. Tsutsui,
			{\it Universal Separability and Entanglement in Identical Particle Systems},
			Phys. Rev. A \textbf{87}, 052313 (2013).
\bibitem{EPUA}		A. P. Balachandran, T. R. Govindarajan, A. R. de Queiroz, and A. F. Reyes-Lega,
			{\it Entanglement and Particle Identity: A Unifying Approach},
			Phys. Rev. Lett. \textbf{110}, 080503 (2013).
\bibitem{CMEIP}		F. Iemini and R. O. Vianna,
			{\it Computable Measures for the Entanglement of Indistinguishable Particles},
			Phys. Rev. A \textbf{87}, 022327 (2013).
\bibitem{QQCFS}		F. Iemini, T. O. Maciel, T. Debarba, and R. O. Vianna,
			{\it Quantifying Quantum Correlations in Fermionic Systems using Witness Operators},
			Quantum Inf. Process. {\bf 12}, 733 (2013).
\bibitem{O14}		M. Oszmaniec,
			{\it Applications of differential geometry and representation theory to description of quantum correlations}
			(PhD Thesis, University of Warsaw, 2014); arXiv:1412.4657 [quant-ph].
\bibitem{QCIP}		F. Iemini, T. Debarba, and R. O. Vianna,
			{\it Quantumness of correlations in indistinguishable particles},
			Phys. Rev. A \textbf{89}, 032324 (2014).
\bibitem{2015}		A. Vald\'es-Hern\'andez, A. P. Majtey, and A. R. Plastino,
			{\it Dynamics of entanglement in systems of identical fermions undergoing decoherence},
			Phys. Rev. A {\bf 91}, 032313 (2015)
\bibitem{KCL05}		J. K. Korbicz, J. I. Cirac, and M. Lewenstein,
			{\it Spin Squeezing Inequalities and Entanglement of N Qubit States},
			Phys. Rev. Lett. {\bf 95}, 120502 (2005); Phys. Rev. Lett. {\bf 95}, 259901(E) (2005).
\bibitem{PS09}		L. Pezz\'e and A. Smerzi,
			{\it Entanglement, Nonlinear Dynamics, and the Heisenberg Limit},
			Phys. Rev. Lett. {\bf 102}, 100401 (2009).
\bibitem{ATSL12}	R. Augusiak, J. Tura, J. Samsonowicz, and M. Lewenstein,
			{\it Entangled symmetric states of N qubits with all positive partial transpositions},
			Phys. Rev. A {\bf 86}, 042316 (2012).
\bibitem{SMLZHPSO14}	P. Hyllus, L. Pezz\'e, A. Smerzi, and G. T\'oth,
			{\it Entanglement and extreme spin squeezing for a fluctuating number of indistinguishable particles},
			Phys. Rev. A {\bf 86}, 012337 (2012);
			H. Strobel, W. Muessel, D. Linnemann, T. Zibold, D. B. Hume, L. Pezz\'e, A. Smerzi, M. K. Oberthaler,
			{\it Fisher information and entanglement of non-Gaussian spin states},
			Science {\bf 345}, 424 (2014).
\bibitem{EEIP}		D. Cavalcanti, L. M. Malard, F. M. Matinaga, M. O. Terra Cunha, and M. Fran\c{c}a Santos,
			{\it Useful entanglement from the Pauli principle},
			Phys. Rev. B {\bf 76}, 113304 (2007).
\bibitem{EEIPlater}	N. Killoran, M. Cramer, and M. B. Plenio,
			{\it Extracting Entanglement from Identical Particles},
			Phys. Rev. Lett. \textbf{112}, 150501 (2014).
\bibitem{MEW}		J. Sperling and W. Vogel, 
			{\it Multipartite Entanglement Witnesses},
			Phys. Rev. Lett. \textbf{111}, 110503 (2013).
\bibitem{Stefan}	S. Gerke, J. Sperling, W. Vogel, Y. Cai, J. Roslund, N. Treps, and C. Fabre,
			{\it Full Multipartite Entanglement of Frequency-Comb Gaussian States},
			Phys. Rev. Lett. {\bf 114}, 050501 (2015).
\bibitem{EMSIP}		J. Grabowski, M. Ku\'{s}, and G. Marmo,
			{\it Entanglement for multipartite systems of indistinguishable particles},
			J. Phys. A: Math. Theor. \textbf{44}, 175302 (2011).
\bibitem{Y08}		K. Yosida,
			{\it Functional Analysis}, 6th ed.
			(Springer, Berlin, 2008), pp. 102; see additionally~\cite{M33}.
\bibitem{M33}		S. Mazur,
			{\it \"Uber konvexe Mengen in linearen normierten R\"aumen},
			Stud. Math. {\bf 4}, 70 (1933).
\bibitem{SV09}		J. Sperling and W. Vogel,
			{\it Necessary and sufficient conditions for bipartite entanglement},
			Phys. Rev. A {\bf 79}, 022318 (2009).
\bibitem{T05}		G. T\'{o}th,
			{\it Entanglement witnesses in spin models},
			Phys. Rev. A {\bf 71}, 010301(R) (2005).
\bibitem{SSV14}		F. Shahandeh, J. Sperling, and W. Vogel,
			{\it Structural Quantification of Entanglement},
			Phys. Rev. Lett. {\bf 113}, 260502 (2014).
\bibitem{HJ13}		R. A. Horn and C. R. Johnson,
			{\it Matrix Analysis}, 2nd ed.
			(Cambridge University Press, Cambridge, UK, 2013), p. 153.
\bibitem{LM77}		J. M. Leinaas and J. Myrheim,
			{\it On the theory of identical particles},
			Il Nuovo Cimento B {\bf 37}, 1 (1977).
\bibitem{W82}		F. Wilczek,
			{\it Quantum Mechanics of Fractional-Spin Particles},
			Phys. Rev. Lett. {\bf 49}, 957 (1982).
\bibitem{DSN} 		J. Sperling and W. Vogel,
			{\it Determination of the Schmidt number},
			Phys. Rev. A \textbf{83}, 042315 (2011).
\bibitem{DNLS}		A. J. Guti\'{e}rrez-Esparza, W. M. Pimenta, B. Marques, A. A. Matoso, J. Sperling, W. Vogel, and S. P\'{a}dua,
			{\it Detection of nonlocal superpositions},
			Phys. Rev. A {\bf 90}, 032328 (2014).
\bibitem{GHZ}		D. M. Greenberger, M. A. Horne, and A. Zeilinger,
			{\it Going Beyond Bells Theorem},
			in {\it Bells Theorem, Quantum Theory, and Conceptions of the Universe}
			(Kluwer Academic, Dordrecht, 1989).

\end{thebibliography}
\end{document}